\documentclass[12pt]{article}
\usepackage{amsfonts,amsmath,amssymb}
\def\L{{\cal{L}}}
\def\bA{{\boldsymbol{A}}}
\def\bD{{\boldsymbol{D}}}
\def\bF{{\boldsymbol{F}}}
 
\def\abc{{\alpha \beta \gamma}}
\def\thetatheta{{{\theta \wedge \theta}}}
\def\Bo{{{(\theta \wedge \theta)_{\it{o}}}}}
\def\Bs{{{(\theta \wedge \theta)_{{\it spin} (7)}}}}
\def\Bg{{{(\theta \wedge \theta)_{{\it g}_2}}}}

\def\g2{{{\it g}_2}}
\def\o{{\it{o}}}
\def\sp{{\it spin}(7)}
\def\Ao{{A_{\it{o}}}}
\def\AS{{A_{{\it spin}(7)}}}
\def\Ag{{A_{{\it g}_2}}}
\def\gt{{\delta}}
\def\w{{\wedge}}

\begin{document}
\title{Chiral gravity in higher dimensions}
\author{Takayoshi Ootsuka~$^{1,}$\thanks{e-mail: ootsuka@cosmos.phys.ocha.ac.jp} ,
Erico Tanaka~$^{1,}$\thanks{e-mail: erico@cosmos.phys.ocha.ac.jp} ,
Kousuke Ura~$^{2,}$\thanks{e-mail: kousuke.ura@hp.com} \\
\quad \\
{\small $^1$ Department of Physics, Ochanomizu University, 
2-1-1 Bunkyo Tokyo, Japan} \\
{\small $^2$ Hewlett-Packard Japan,Ltd. Crystal Tower, 1-2-27,
Shiromi Chuo Osaka, Japan}}
\date{}
\maketitle

\begin{abstract}
We construct a chiral theory of gravity in 7 and 8 dimensions, which are
equivalent to Einstein-Cartan theory using less variables.
In these dimensions, we can construct such higher dimensional chiral gravity because of the existence of gravitational instanton.
The octonionic-valued variables in the theory represent the deviation from 
the gravitational instanton, and from their non-associativity, prevents the theory to be $SO(n)$ gauge invariant.
Still the chiral gravity holds $G_2$ (7-D), and $Spin(7)$ (8-D) gauge symmetry.

\end{abstract}



\section{Introduction}

Recently, several researches were made for $G_2$ and $Spin(7)$ holonomy
manifold in 7 and 8 dimensions~\cite{Joyce,C-G-L-P,K-Y,N-R}.
Such manifolds have reduced holonomy, i.e., 
in 7 dimension $SO(7)$ holonomy reduced to $G_2$, 
in 8 dimension $SO(8)$ to $Spin(7)$. Also, from these properties, 
it is known that there exists covariantly constant spinor on these 
manifolds, and therefore they have been a plausible candidate for 
compactification of extra dimensions which shows up in M-theory and 
superstring theory~\cite{Acharya,A-W,C-G-L-P2}. 
For the meanwhile, 4-dimensional quantum loop gravity revealed some 
non-perturbative effects of gravity, making use of Ashtekar 
formalism~\cite{Ashtekar,Rovelli,Smolin}. 
The Ashtekar formalism used in this case is a 4-dimensional chiral 
theory~\cite{C-D-J-M}, 
which is a theory presented in partial variables (giving the 
naming of chiral) of Einstein-Cartan theory.
Namely, original spin connection $SO(4)$ or $SO(3,1)$ are decomposed 
and the half $SO(3)$ are used in chiral theory. 
One specific point of 4-dimensional chiral gravity, 
is that by setting the connection to zero for the chiral equation of 
motion, the 4-dimensional special holonomy manifold, 
$Sp(1)$ holonomy manifold (in other words, HyperK\"ahler manifold;
or gravitational instanton) could be derived 
straight-forward~\cite{Robinson,O-M-Y-Z}.
It might be said that 4-dimensional chiral gravity states the difference 
from the gravitational instanton. 
This is in contrast with Einstein-Cartan theory, whose non-zero connection describes the deviation from flat space-time.
In the following discussion, we show in certain higher dimensions, 
there exists a chiral formalism of gravity. That is, when special 
holonomy manifold exists, instead of Einstein-Cartan's $SO(n)$ 
connection, we could use partial ("chiral") variables and 
construct a theory which is equivalent to Einstein-Cartan theory. 
However, in higher dimensions, chiral variables could not be understood
as connections as in 4 dimension. Still it has a local gauge symmetry
of $G_2$ (7-D) and $Spin(7)$ (8-D), which is related to 
non-associative algebra, octonions.

\section{Einstein Cartan theory}

We assume that the following discussions are all made in Euclidean. 
 
We start from the Euclid Einstein-Cartan Lagrangian,
 which form is valid in any $n$ dimensions,
\begin{eqnarray}
\L_{E.C.}(A,\theta) = \left< F\w *(\theta \w \theta)\right>  \label{e.c.},
\end{eqnarray}
where $A$ is the $so(n)$ connection, $\theta$ the $n$-bein 1-form, 
and $F$ a $so(n)$ curvature. Their explicit definition are given by,  
 \begin{eqnarray}
    &&A = \frac12 A^{ab} \otimes M_{ab},\\
    &&F = dA + \frac12 [A \w A]\, \nonumber \\
    &&\theta = \theta^a \otimes P_a, 
    \quad \theta \w \theta = \frac12 \theta^a \w \theta^b 
    \otimes M_{ab}\quad \nonumber \\
    &&*(\thetatheta) = 
    \frac12 \frac{1}{(n-2)!}\epsilon_{abc_1 \cdots c_{n-2}}
    \theta^{c_1 \cdots c_{n-2}} \otimes M_{ab}.
  \end{eqnarray}
Above, $\theta^{c_1 \cdots c_{n-2}}$ is the abbreviation of $\theta^{c_1} \w\cdots  \w \theta^{c_{n-2}}$, and from now on we use this useful notation. 
Generators $M_{ab}$ and $P_c$ form the Poincar\'e algebra shown below,
  \begin{eqnarray}
    &&[M_{ab},M_{cd}]=\delta_{ac}M_{bd}+\delta_{bd}M_{ac}
    -\delta_{ad}M_{bc}-\delta_{bc}M_{ad} \nonumber \\
    &&[M_{ab},P_c]=\delta_{ac}P_b - \delta_{bc}P_a, \quad
    [P_a,P_b]=0 .
  \end{eqnarray}
There also is a property of killing forms, which we will use 
frequently in the calculations.
\begin{eqnarray}
 &&\left< P_a , P_b \right> = \delta_{ab}, \nonumber \\
 &&\left< M_{ab} , M_{cd} \right> = \delta_{ac}\delta_{bd}
   - \delta_{ad}\delta_{bc}, \nonumber \\ 
 &&  \left< [M_{ab} ,P_c] ,P_d \right> + \left< P_c ,
 [M_{ab},P_d] \right> = 0, \nonumber \\
 && \left< [M_{ab},M_{cd}] ,M_{ef} \right> 
 + \left< M_{cd},[M_{ab},M_{ef}] \right> = 0 .
\end{eqnarray}
The equation of motion derived from this Lagrangian is,
\begin{eqnarray}
&&D*(\thetatheta) = 0,\nonumber \\
&&\epsilon_{abc_1 \cdots c_{n-2}}F^{ab} \w \theta^{c_1 \cdots c_{n-3}} = 0.
\end{eqnarray}
The chiral theory we will construct is equivalent to this 
Einstein-Cartan (E.C.) theory in a certain aspect.

\section{4-dimensional chiral gravity}

We begin by reviewing the 4-dimensional chiral gravity~\cite{Giulini}, 
which will be useful when making comparison to higher dimensions. 
The well-known decomposition of $so(4)$ and projection operators are 
\begin{eqnarray}
&&so(4) = so(3)_+ \oplus so(3)_- \nonumber \\
&&{P_+}_{\mu \nu \rho \sigma} = \frac14 \left( \delta_{\mu \rho} \delta_{\nu \sigma} - \delta_{\mu \sigma} \delta_{\nu \rho} + \epsilon_{\mu \nu \rho \sigma} \right),\nonumber\\
&&{P_-}_{\mu \nu \rho \sigma} = \frac14 \left( \delta_{\mu \rho} \delta_{\nu \sigma} - \delta_{\mu \sigma} \delta_{\nu \rho} - \epsilon_{\mu \nu \rho \sigma} \right),
\end{eqnarray}
the Greek subscript runs from 0 to 3. 
Applying the decomposition, we may write the connection $A$ and $(\thetatheta)$,
\begin{eqnarray}
&&A = A_+ + A_-,\nonumber\\
&&(\thetatheta) = (\thetatheta)_+ + (\thetatheta)_-, \nonumber
\end{eqnarray}
and similarly,
\begin{eqnarray} 
F_+ = dA_+ +[A_+ \w A_+],\nonumber\\
F_- = dA_- +[A_- \w A_-].
\end{eqnarray}
$A_\pm$ are the $so(3)_\pm$ connection, and $F_\pm$ the $so(3)_\pm$ curvature.
The fact that $so(4)$ decompose to $so(3)_\pm$ as a Lie algebra is a crucial point in 4 dimension. 
It is easy to see that,
\begin{eqnarray}
*(\thetatheta)_+ = +(\thetatheta)_+, \nonumber \\ 
*(\thetatheta)_- = -(\thetatheta)_-, 
\end{eqnarray}
so the Lagrangian in these terms become
\begin{eqnarray}
\L_{4E.C.} = \left<  F_+ \w (\thetatheta)_+ \right>  - \left<  F_- \w (\thetatheta)_- \right> .
\end{eqnarray}
To make this into chiral theory, one needs to delete half of the 
connection, $A_-$. One may recall that Einstein-Cartan gravity gives 
torsion-less condition for its equation of motion, $D\theta = 0$. 
When one solves this equation for the connection $A$ and restore the result
 back to the Lagrangian, one merely gets the Einstein-Hilbert Lagrangian.
Instead of solving the equation for all $A$, if one solves 
for $A_-$ only and substitute the result to the Lagrangian, one would get 
a Lagrangian written in $A_+$. To solve for partial variables, and replace 
the original variables with the remaining variables, 
is the main idea of chiral gravity.
Here for some calculation simplicity, we achieve the same result by 
just adding an exact term to the above Lagrangian.
Specifically, 
\begin{eqnarray}
&&\L_{4E.C.} + d\left<  T \w \theta \right>  \nonumber\\
&&= \L_{4E.C.} + \left<  F \w (\thetatheta) \right>  + \left<  T \w T \right>  \nonumber \\
&&=: \L_{4chiral} + \left<  T \w T \right> . 
\end{eqnarray}
$T$ being the torsion $T = D\theta$.  
So, 
\begin{eqnarray}
\L_{4E.C.} \equiv \L_{4chiral} \quad ({\it{mod}}.\, \, T = 0 ),
\end{eqnarray}
in a sense that both Lagrangian gives the same solution for Einstein equation.
The description of the chiral Lagrangian is, 
\begin{eqnarray}
\L_{4chiral}(\bA,\theta) &=& \left<  \left( d\bA + \frac12 [\bA \w \bA] \right) \w (\thetatheta)_+ \right> , 
\end{eqnarray}
where $A_+$ is replaced by $\bA$, 
and the equation of motion derived from this Lagrangian is, 
\begin{eqnarray}
&&\bD(\thetatheta)_+ = 0, \nonumber \\ \quad
&&[\bF \w \theta] = 0,
\end{eqnarray}
boldface denoting they are written in $\bA$.
Chiral Lagrangian is the preliminary form for the 
Ashtekar theory of gravity~\cite{C-D-J-M}, 
which we hope to find for the higher 
dimensional case as well. Its chiral construction will be given in 
the following sections.

\section{Octonions} \label{sec.octonion}

Before going into the construction of the theory, let us briefly give some outlines of octonions($\mathbb{O}$)~\cite{Tze,Baez}.
This peculiar algebra appears when one decomposes $so(8)$ and $so(7)$ considering the particular solutions of Einstein equations, 
the gravitational instantons~\cite{Bakas}. 
Octonions are the biggest alternative division algebra, which are non-commutative and also non-associative. 
Their basis satisfies the following relations,
\begin{eqnarray}
   &&a \in \mathbb{O} \quad  a = a_0e_0+a_1e_1+ \cdots +a_7e_7 \quad 
	 a_0\cdots a_7 \in \mathbb{R} \nonumber \\ \quad
	 &&e_0:=1, \quad 
	 e_ie_j := -\delta_{ij} +\varphi_{ijk}e_k, \quad (i,j = 1 \cdots 7) \nonumber\\
   &&  \varphi_{abc} = 1 \quad \mbox{if} \quad 
     (abc)=\left\{
       \begin{array}{l}
         (123),(516),(624),(435),\nonumber\\
         (471),(673),(572)
       \end{array}\right. \nonumber \quad \\ 
   &&\varphi_{abc} = 0 \quad \mbox{else} \nonumber\\
   && e_a(e_b\,e_c)-(e_a\,e_b)e_c = \psi_{abcd}e_d, \quad 
   \psi_{abcd} = -\frac1{3!} \epsilon_{abcdefg} \varphi_{efg}. 
 \end{eqnarray}
They are similar to quarternions except the last line, 
claiming non-associativity. The structure constant $\varphi_{abc}$ represents non-commutativity, and $\psi_{abcd}$ represents non-associativity. 
Bellow are some useful relations of these structure constants that we will make use of in the later calculations.
\begin{eqnarray}
	& &\varphi_{abe}\varphi_{cde} = \delta_{ac}\delta_{bd} - 
	\delta_{ad}\delta_{bc} + \psi_{abcd} ,\nonumber\\
	& &\varphi_{ade}\psi_{bcde} = 4 \varphi_{abc} ,\nonumber\\
	& &\varphi_{abf}\psi_{cdef} = \frac12 {\varphi^{[a}}_{[cd}\delta^{b]}_{e]}  ,\nonumber\\
  & &\psi_{abij}\psi_{cdij} = 4 \left( \delta_{ac}\delta_{bd} - 
	\delta_{ad}\delta_{bc} \right) + 2 \psi_{abcd} , \nonumber\\
  & &\psi_{abci}\psi_{defi} =  \delta^{[a}_{[d}\delta^{b}_{e} \delta^{c]}_{f]} - \varphi_{abc}\varphi_{def} 
  + \frac14 {\psi^{[ab}}_{[de}\delta^{c]}_{f]}.
\end{eqnarray}
Square brackets including indices denote anti-symmetrization without 
factor $1/n!$.
We also introduce some symbols which is convenient in the case for 8 dimension. 
\begin{eqnarray}
&&\zeta^a_{\mu \nu} = -\zeta^a_{\nu \mu}, \quad 
\zeta^a_{0 b} = \delta^a_{b}, \quad 
\zeta^a_{bc} = -\varphi_{abc} \nonumber \\ 
&&\Lambda^{ab}_{\mu \nu} = -\Lambda^{ab}_{\nu \mu}, \quad
\Lambda^{ab}_{\mu \nu} = -\Lambda^{ba}_{\mu \nu}, \nonumber \\
&&\Lambda^{ab}_{0 c} = -\varphi_{abc}, \quad 
\Lambda^{ab}_{cd} = \delta^a_c\delta^b_d - \delta^b_c\delta^a_d - \psi_{abcd}. 
\end{eqnarray}
The Latin subscript runs from 1 to 7, and the Greek from 0 to 7.
The first symbol $\zeta^a_{\mu \nu}$ is similar to the t'Hooft symbol in 4 dimension.

The chiral construction in 8 and 7 dimensions are deeply related to these octonion properties.

\section{8-dimensional chiral gravity}

The decomposition of $so(8)$ we chose is 
\begin{eqnarray}
so(8) = {\it o} \oplus {\sp}  ,
\end{eqnarray}
which is a natural splitting when considering particular solutions of Einstein equation, i.e., gravitational instantons, in 8 dimension~\cite{Bakas}.
This decomposition is $Spin(7)$ invariant, and $\o$ is isomorphic to the 
imaginary part of octonions. The algebra for this decomposition are,
\begin{eqnarray}
&&[\sp,\sp] \subset \sp, \nonumber \\ \quad
&&[\sp,{\o}] \subset {\o}, \nonumber \\ \quad 
&&[{\o},{\o}] \subset \sp. \label{octbra1}
\end{eqnarray}
The projection operators are
 \begin{eqnarray}
 &&{P_{\it o}}_{\mu \nu \rho \sigma} = \frac18 \{ \delta_{\mu \rho}\delta_{\nu \sigma} - \delta_{\mu \sigma}\delta_{\nu \rho} + f_{\mu \nu \rho \sigma} \} , \nonumber\\
 &&{P_{\sp}}_{\mu \nu \rho \sigma} = \frac18 \{ 3(\delta_{\mu \rho}\delta_{\nu \sigma} - \delta_{\mu \sigma}\delta_{\nu \rho}) - f_{\mu \nu \rho \sigma} \} , 
\end{eqnarray}
where Greek subscript runs from 0 to 7.
The structure constant $f_{\mu \nu \rho \sigma}$ satisfies the self-duality condition,
\begin{eqnarray}
f_{\mu \nu \rho \sigma} = \frac{1}{4!}\epsilon_{\mu \nu \rho \sigma \tau \lambda \eta \zeta}f_{ \tau \lambda \eta \zeta}
\end{eqnarray}
and could be written in terms of the octonion structure constant,   
\begin{eqnarray}
f_{abcd} = \psi_{abcd}, \quad 
f_{0abc} = -\varphi_{abc}. 
\end{eqnarray}
The Latin subscripts run from 1 to 7.
The useful forms of the projection operators are 
\begin{eqnarray}
{P_{\it o}}_{\mu \nu \rho \sigma} = \frac18 \zeta^a_{\mu \nu} \zeta^a_{\rho \sigma}, \nonumber \quad 
{P_{\sp}}_{\mu \nu \rho \sigma} = \frac{1}{16}\Lambda^{ab}_{\mu \nu} \Lambda^{ab}_{\rho \sigma}.
\end{eqnarray}
Same index means taking summation.
In terms of these operators, we change the generator to $E_a$ for $\o$, and $S_{ab}$ for $\sp$,
\begin{eqnarray}
E_a = \frac14 \zeta^a_{\mu \nu} M_{\mu \nu}, \quad 
S_{ab} = \frac 14 \Lambda^{ab}_{\mu \nu} M_{\mu \nu}.
\end{eqnarray}
They satisfy the following algebra and killing form,
\begin{eqnarray}
&&[ E_a, E_b ] = S_{ab}, \quad
[ E_a, S_{bc} ] = -\left( \delta_{ab} \delta_{ck} - \delta_{ac} \delta_{bk} \right) E_k, \quad \nonumber\\
&&[S_{ab},S_{cd}]=\delta_{ac}S_{bd}+\delta_{bd}S_{ac}
    -\delta_{ad}S_{bc}-\delta_{bc}S_{ad} \nonumber \\
&&\left< E_a ,E_b \right> = \delta_{ab}, \quad 
\left< S_{ab} , S_{cd} \right> = \delta_{ac}\delta_{bd}
   - \delta_{ad}\delta_{bc}, \nonumber \\ 
&&\left< [E_a, E_b], S_{cd} \right> + \left< E_a, [S_{cd}, E_b] \right> = 0, \nonumber\\
&&\left< E_a, S_{bc} \right> = 0. \label{kil1}
\end{eqnarray}
In accordance with 4-dimensional case, we start from the Einstein-Cartan Lagrangian (\ref{e.c.}). 
Applying the projection to connection $A$ and $\thetatheta$,  
\begin{eqnarray}
&&A = \Ao + \AS,\nonumber \\
&&\thetatheta = (\thetatheta)_{\o} + (\thetatheta)_{\sp},
\end{eqnarray}
we will construct the 8-dimensional chiral theory with $\Ao$ only.
Performing some calculations, one could find the following relations,
\begin{eqnarray}
&&*\Bo = \frac13 \Bo \w \Psi, \nonumber\\
&&*\Bs = - \Bs \w \Psi,
\end{eqnarray}
and $\Psi$ is a Cayley 4-form, which is given as 
\begin{eqnarray}
\Psi = \frac{1}{4!}f_{\mu \nu \rho \sigma}\theta^{\mu \nu \rho \sigma}. 
\end{eqnarray}
However, since the decomposition (\ref{octbra1}) is not a decomposition 
as a Lie algebra, the curvature component contains both $\Ao$ and $\AS$,
\begin{eqnarray}
&&F_{\o} = d\Ao + [\Ao \w \AS],\nonumber\\
&&F_{\sp} = d\AS + \frac12 [\Ao \w \Ao] + \frac12 [\AS \w \AS].
\end{eqnarray}
Lagrangian now could be written,
\begin{eqnarray}
\L_{8E.C.} = \frac13 \left<  F_{\o} \w \Bo \right>  \w \Psi
- \left<  F_{\sp} \w \Bs \right>  \w \Psi. 
\end{eqnarray}
And we could apply the same method of adding an exact term to the Lagrangian as in 4 dimension, 
\begin{eqnarray}
\hspace{-0.5cm} &&\L_{8E.C.} + d\left(\left< T \w \theta \right> \w \Psi\right) \nonumber \\
\hspace{-0.5cm} &&= \L_{8E.C.} + \left<  F \w (\thetatheta) \right>  \w \Psi 
 + \left<  T\w T \right>  \w \Psi - \left<  T \w \theta \right>  \w d \Psi \nonumber \\
\hspace{-0.5cm} &&= \frac43 \left<  F_{\o} \w (\thetatheta)_{\o} \right>  \w \Psi + \left<  T\w T \right>   \w \Psi - \left<  T \w \theta \right>  \w d\Psi, \label{pre8ch} 
\end{eqnarray}
where $T=D\theta$ is the torsion. 
Recalling that $F_o$ contains the variable $\AS$, one needs to remove this from (\ref{pre8ch}).
We work this out by using the killing form relation and the torsion-less condition, 
\begin{eqnarray}
&& D(\thetatheta) = 2 \, T \w \theta = 0, \nonumber \\
&& \Leftrightarrow  
\left\{
	\begin{array}{l}
		d\Bo + [\Ao \w \Bs] + [\AS \w \Bo]  = 0 \\
		d\Bs + [\Ao \w \Bo] + [\AS \w \Bs] = 0 \label{DBo}
	\end{array} \right. .
\end{eqnarray}
Leaving out the terms including $T$, (\ref{pre8ch}) becomes,
\begin{eqnarray}
&& \left< \left( d\Ao + [ \Ao \w \AS ] \right) \w \Bo \right> \w \Psi  \nonumber \\
&& = \left\{ \left< d\Ao \w \Bo \right> 
+ \left< \Ao \w [ \AS \w \Bo ] \right> \right\} \w \Psi \nonumber \\
&& = \left\{ \left< d\Ao \w \Bo \right> 
+ \left< \Ao \w \left( -d\Bo - [ \Ao \w \Bs ] \right) \right> \right\} \w \Psi \nonumber \\
&& = \left\{ d \left< \Ao \w \Bo \right> 
- \left< [ \Ao \w \Ao ] \w \Bs \right> \right\} \w \Psi, 
\end{eqnarray}
the first and third equality using the property of killing forms,
and the second equal using the first of (\ref{DBo}). 
We define   
\begin{eqnarray}
\hspace{-0.5cm}\L_{8chiral}(\bA,\theta) 
:= \left\{ d \left<  \bA \w \Bo \right>  
- \left<  [\bA \w \bA] \w \Bs \right>  \right\} \w \Psi,
\end{eqnarray}
where $\Ao$ is replaced by $\bA$. 
This is the chiral Lagrangian, in a sense that is 
\begin{eqnarray}
\L_{8E.C.} \equiv \L_{8chiral} \quad (\it{mod}. \,\, T = 0 ) .
\end{eqnarray}
However, variable $\bA$ could no longer be thought as a connection.

To check the validity of this Lagrangian, we derive the equation of motion 
from this Lagrangian and verify that it gives the particular solutions of 
Einstein equation, gravitational instanton. 
The equation of motion is,
\begin{eqnarray}
&&\Bo \w d\Psi -2 [\bA \w \Bs ] \w \Psi = 0, \label{8Tnull} \\
&&\left\{ d\left<  \bA \w \Bo \right>  
- \left<  [\bA \w \bA] \w \Bs \right>  \right\} 
\w \left( \frac{1}{3!} f_{\abc \lambda} \theta^\abc \right)  \nonumber \\
&&- \frac12 \left\{ \zeta^a_{\mu \lambda}\bA^a \w \theta^\mu \w d\Psi 
+ \Lambda^{ab}_{\mu \lambda} \bA^a \w \bA^b \w \theta^\mu \w \Psi \right\} = 0.
\label{8chE} 
\end{eqnarray}
The former corresponds to torsion-less condition, and the latter to Einstein equation.
Instanton solution is obtained with the condition $\bA = 0$.
Second equation becomes trivial, and it can be shown (see Appendix) that the equation 
(\ref{8Tnull}),
\begin{eqnarray}
(\thetatheta)_{\o} \w d \Psi = 0 \label{8Tnull2}
\end{eqnarray}
is equivalent to 
\begin{eqnarray}
d\Psi = 0.
\end{eqnarray}
It is known that the $Spin(7)$ holonomy manifold, which is a 
gravitational instanton of 8 dimension, could be constructed from the 
$Spin(7)$ invariant self-dual 4-form which satisfies the above condition. 

Finally, we briefly show the Lagrangian has $Spin(7)$ gauge symmetry.
The $Spin(7)$ gauge transformation is given as,
\begin{eqnarray}
&&\gt \bA = [ \chi, \bA ], \nonumber \\
&&\gt \Bo = [ \chi, \Bo ] \nonumber,  \\
&&\gt \Bs = [ \chi, \Bs ].
\end{eqnarray}
Where $\chi$ is $spin(7)$-valued gauge function.
The 4-form $\Psi$ could be written as 
\begin{eqnarray}
\Psi = \frac23 \left< \Bo \w \Bo \right>,
\end{eqnarray}
so it is $Spin(7)$ gauge invariant,
\begin{eqnarray}
\gt \Psi = \frac43 \left< [\chi, \Bo] , \Bo \right> 
= \frac43 \left< \chi, [ \Bo, \Bo ]  \right> = 0.
\end{eqnarray}
The Lagrangian transforms similarly,  
\begin{eqnarray}
&&\gt \L_{8chiral}(\bA,\theta) 
= \left\{ d \left<  [\chi,\bA] \w \Bo \right> 
+ d \left< \bA \w [\chi, \Bo ] \right> \right. \nonumber \\ 
&&+  2 \left< [ \chi, \bA ],[ \bA \w \Bs ] \right> 
+ \left< [\bA \w \bA] ,[\chi, \Bs] \right>  \nonumber \\
&& \left. - \left<  [\bA \w \bA] \w \Bs \right>  \right\} \w \Psi = 0,
\end{eqnarray}
making use of killing form relation and Jacobi identity.

\section{7-dimensional chiral gravity}

The appropriate decomposition for $so(7)$ and the algebra for them are
\begin{eqnarray}
&& so(7) = {\o} \oplus \g2, \nonumber \\
&& [ \g2, \g2 ] \subset \g2 , \nonumber \\
&& [ \o, \g2 ] \subset \o, \nonumber \\
&& [ \o, \o ] \subset so(7) \label{octbra2}. 
\end{eqnarray}
and its projection operators are
\begin{eqnarray}   
   && {P_{\o}}_{abcd} = \frac16 \{ (\delta_{ac} \delta_{bd} - \delta_{ad} \delta_{bc})
       + \psi_{abcd}\}, \nonumber\\
 && {P_{\g2}}_{abcd} = \frac13 \{ (\delta_{ac} \delta_{bd} - \delta_{ad} \delta_{bc}) 
       - \frac12 \psi_{abcd}\}, 
\end{eqnarray}
Latin subscripts running from 1 to 7. 
$\psi_{abcd}$ and $\varphi_{abc}$ are the octonion structure 
constants introduced in section \ref{sec.octonion}.
Only $P_{\o}$ could be represented by a quadratic form this time,
\begin{eqnarray}
{P_{\o}}_{abcd} = \frac16 \varphi_{abi}\varphi_{cdi}.
\end{eqnarray} 
We transform the generators to $e_a$ for $\o$ and $g_{ab}$ for $\g2$,
\begin{eqnarray}
e_a = \frac12 \varphi_{aij}M_{ij}, \quad 
g_{ab} = \frac13 \left(2M_{ab} - \frac12 \psi_{abij} M_{ij} \right) .
\end{eqnarray}
They satisfy the following algebra and killing forms,
\begin{eqnarray}
&& [ e_a, e_b ] = 3g_{ab} - \varphi_{abc}e_c, \quad
[ g_{ab}, e_c ] = \frac23\left\{ \delta_{ac} \delta_{bk} - \delta_{ak} \delta_{bc} -\frac12 \psi_{abck} \right\} e_k, \quad \nonumber \\
&& \left< e_a ,e_b \right> = 3\delta_{ab}, \quad 
\left< g_{ab} , g_{cd} \right> = \frac23 \left\{ \delta_{ac}\delta_{bd}
   - \delta_{ad}\delta_{bc} -\frac12 \psi_{abcd} \right\}, \nonumber \\ 
&& \left< [e_a, e_b], g_{cd} \right> + \left< e_a, [g_{cd}, e_b] \right> = 0, \nonumber\\
&& \left< e_a, g_{bc} \right> = 0.
\end{eqnarray}
As in previous section, we start from the Einstein-Cartan 
Lagrangian (\ref{e.c.}).
Applying the projection to connection $A$ and $\thetatheta$,  
\begin{eqnarray}
&&A = \Ao + \Ag, \nonumber\\
&&\thetatheta = \Bo + \Bg,
\end{eqnarray}
we construct the 7-dimensional chiral theory with $\Ao$ only.
Performing some calculations, one could find the following relations,
\begin{eqnarray}
&&*\Bo = \frac12 \Bo \w \varphi, \nonumber\\
&&*\Bg = - \Bg \w \varphi,
\end{eqnarray}
where $\varphi$ is a Cayley 3-form, given as 
\begin{eqnarray}
\varphi = \frac{1}{3!}\varphi_{abc}\theta^{abc}. 
\end{eqnarray}
Now, by simple calculation we have the following identity, 
\begin{eqnarray}
(\thetatheta)_{\o}\w \varphi = \frac23 \theta \w \psi,
\end{eqnarray}
where $\theta$ is no longer $\theta^a \otimes P_a$, but an adjoint variable,
$\theta = \theta^a \otimes e_a$, and  
$\psi$ is a Cayley 4-form, which is given as 
\begin{eqnarray}
\psi = \frac{1}{4!}\psi_{abcd}\theta^{abcd}. 
\end{eqnarray}
So we can write
\begin{eqnarray}
*(\thetatheta) = \frac13 \theta \w \psi - \Bg \w \varphi. 
\end{eqnarray}
The decomposed curvature component contains both the $\Ao$ and $\Ag$,
\begin{eqnarray}
&&F_{\o} = d\Ao + \frac12[\Ao \w \Ao]_\o + [\Ao \w \Ag],\nonumber\\
&&F_{\g2} = d\Ag + \frac12 [\Ao \w \Ao]_\g2 + \frac12 [\Ag \w \Ag].
\end{eqnarray}
The Einstein-Cartan Lagrangian becomes,
\begin{eqnarray}
\L_{7E.C.} = \frac13 \left<  F_{\o} \w \theta \right>  \w \psi 
- \left<  F_{\g2} \w (\thetatheta)_{\g2} \right>  \w \varphi.
\end{eqnarray}
By the same method used in 8 dimension, we delete $\Ag$ from this 
Lagrangian. Namely, first add an exact term and modify the Lagrangian in a 
convenient form, 
\begin{eqnarray}
\hspace{-0.5cm}&&\L_{7E.C.} + d\left(\left< T \w \theta \right> \w \varphi\right) \nonumber \\
\hspace{-0.5cm} &&= \L_{7E.C.} + \left<  F \w (\thetatheta) \right>  \w \varphi 
 + \left<  T\w T \right>  \w \varphi - \left<  T \w \theta \right>  \w d\varphi \nonumber \\
\hspace{-0.5cm} &&= \frac13 \left<  F_{\o} \w \theta \right>  \w \psi 
+ \left<  F_{\o} \w \Bo \right>  \w \varphi 
+ \left<  T\w T \right>   \w \varphi 
- \left<  T \w \theta \right>  \w d\varphi, \label{pre7ch} 
\end{eqnarray}
then next delete the remaining $\Ag$ in $F_{\o}$ by the condition $D(\thetatheta) = 0 $, and the property of killing forms. 
\begin{eqnarray}
&& D(\thetatheta) = 2 \, T \w \theta = 0, \\
&& \Leftrightarrow  
\left\{
	\begin{array}{l}
		d(\thetatheta)_{\o} + [\Ao \w (\thetatheta)_{\o}]_{\o} + [\Ao \w (\thetatheta)_{\g2}] + [\Ag \w (\thetatheta)_{\o}]  = 0 \nonumber\\
		 d(\thetatheta)_{\g2} + [\Ao \w (\thetatheta)_{\o}]_{\g2} + [\Ag \w (\thetatheta)_{g2}] = 0 \nonumber
	\end{array} \right.
\end{eqnarray}
The modified Lagrangian, (the first and second terms of (\ref{pre7ch})) 
is a natural form because it also could be derived from 8-dimensional 
Lagrangian by reduction to 7 dimension. The way of adding two forms of 
essentially same terms may be called as a tuning,  
and though there are various way to add exact term to modify the Lagrangian, 
the particular way we chose was a necessary procedure to give gravitational instantons at $\bA = 0$.
The chiral Lagrangian obtained is:
\begin{eqnarray}
\L_{7chiral}(\bA,\theta) &&=
 \frac13 \left\{ d \left<  \bA \w \theta \right> - \frac12\left<  [\bA \w \bA] \w \theta \right>  \right\} \w \psi \nonumber\\
	 &&\begin{array}{l}
		 + \left\{ d \left<  \bA \w (\thetatheta) \right>  
		 - \frac12 \left<  [\bA \w \bA] \w (\thetatheta)_{\o} \right>  \right. \\
		 - \left. \left<  [\bA \w \bA] \w (\thetatheta)_{\g2} \right>  \right\} \w \varphi,
	 \end{array}
\end{eqnarray}
which satisfy      
\begin{eqnarray}
\L_{7E.C.} \equiv \L_{7chiral} \quad ({\it mod}.\, \, T = 0 ).
\end{eqnarray}
Now check the equation of motion below includes the gravitational instanton,
\begin{eqnarray} 
&&- \frac13 \theta \w \psi + (\thetatheta)_{\o}\w d\varphi \nonumber\\
&&- \frac13 [\bA \w \theta]_{\it o} \w \psi
+  [\bA \w \Bo ]_{\it o} \w \varphi 
- 2 [\bA \w (\thetatheta)_{\g2} ] \w \varphi = 0, \label{7Tnull} \\
&&\left\{ d\left< \bA \w \Bo \right> + \left< [\bA \w \bA]_\g2 \w \Bg 
\right> \right\} \w \left( \frac12 \varphi_{aij} \theta^{ij} \right) \nonumber \\
&&-\left\{ d\left< \bA \w \theta \right> 
- \frac12 \left< [\bA \w \bA]_{\o} \w \theta \right> \right\} \w 
\left( \frac{1}{3!} \psi_{aijk} \theta^{ijk} \right) \nonumber \\
&&+ \bA^a \w d\psi + \frac32 \varphi_{aij}\bA^i \w \bA^j \w \psi \nonumber \\
&&+ \varphi_{aij}\bA^i \w \theta^j \w d\varphi 
- \frac83 \bA^a \w \bA^i \w \theta^i \varphi 
+ \frac23 \psi_{aijk} \bA^i \w \bA^j \w \theta^k \varphi = 0. \label{7chE} 
\end{eqnarray}
Here $\Ao$ is replaced by $\bA$.
Instanton solution is obtained with the condition $\bA = 0$. 
Again, (\ref{7chE}) becomes trivial, and by rather tedious calculations (see Appendix), 
it can be shown that the equation
\begin{eqnarray}
-\frac13 \theta \w d \psi + (\thetatheta)_{\o} \w d \varphi = 0 \label{7Tnull2}
\end{eqnarray}
is equivalent to:
\begin{eqnarray}
d\psi = 0, \quad d\varphi = 0. 
\end{eqnarray}
The $G_2$ manifold, the gravitational instanton of 7 dimension, is constructed from the closed Cayley 3-form and 4-form, $\varphi$ and $\psi$.

It could be verified that Lagrangian has $G_2$ gauge symmetry. 
The $G_2$ gauge transformation is given as,
\begin{eqnarray}
&&\gt \bA = [ \chi, \bA ], \nonumber \\
&&\gt \Bo = [ \chi, \Bo ] \nonumber,  \\
&&\gt \Bg = [ \chi, \Bg ],
\end{eqnarray}
where $\chi$ is $\g2$-valued gauge function.
$\psi$ and $\varphi$ could be written as 
\begin{eqnarray}
\psi = \frac12 \left< \Bo \w \Bo \right>, \quad
\varphi = \frac13 \left< \theta \w \Bo \right>,
\end{eqnarray}
and are clearly $G_2$ gauge invariant.
By using same relation of killing forms and Jacobi identity as in 8 dimension, it is easy to verify  
\begin{eqnarray}
&&\gt \L_{7chiral}(\bA,\theta)  = 0 .
\end{eqnarray}

\section{Discussion}

We presented the chiral theory of gravity in 7 and 8 dimensions, which are described by less variables than original Einstein-Cartan theory. These chiral variables $\bA$ also indicates the deviation from the gravitational instantons.
However, in these higher dimensions, the chiral variables could not be thought as a connection in comparison to 4-dimensional case. Namely, it does not transform as the usual connection by gauge transformation. It is verified that these chiral theory has $G_2$ and $Spin(7)$ gauge invariance while they lose $SO(7)$ and $SO(8)$ gauge symmetry.
This corresponds to the fact that the chiral variable $\bA$ must be regarded as the non-associative algebra, octonionic-valued variable. Furthermore, again in contrast to 4 dimension, the chiral Lagrangian is not in the form of $BF$ type, especially the term corresponding to curvature $F$ deviates, and this also comes from the octonionic properties of the chiral variables. 
It is under question that whether like the $BF$ theory this Lagrangian would be quantizable in this form, but the prospects are that it could be utilized for the higher dimensional quantum gravity as the Ashtekar theory. As the 
7 and 8-dimensional chiral gravity reflects the geometric relation of $G_2$ and $Spin(7)$ manifold, it is hopeful that similar relation as in the case of 3 and 4 dimensions exist and enables one to construct higher dimensional Ashtekar gravity.
It is also hoped for the discussion of compactification of 7 or 8 dimensions, as in the Kaluza-Klein scenario of M-theory or 11-dimensional supergravity. In the course of these discussions, we also think it is significant to generalize the theory with supersymmetry. The construction of self-dual supergravity with reduced $Spin(7)$ and $G_2$ holonomy in 8 and 7 dimensions respectively, are already embodied~\cite{N-R}, so one can verify the formulated theory when it is achieved.
Finally, in higher dimensions where the special holonomy manifold exists such as $4n\, (n=1,2,\cdots)$ or $2n \, (n=2,3,\cdots)$ dimension; namely, $Sp(n)$ holonomy manifold or HyperK\"ahler manifold in $4n$ dimension, and $SU(n)$ holonomy manifold or special K\"ahler manifold in $2n$ dimension; such chiral gravity should be constructed, and it is now under preparation.
As in the dimensions 4, 7 and 8, their chiral variables should represent the deviation from these gravitational instantons. That is, there must be careful tuning to make the chiral variable $\bA$ represent the projected part of connection of the original E.C. theory. For instance, in 7 dimension, construction of the chiral action was performed that its chiral variable $\bA$ would be the octonionic part of the original $SO(7)$ connection.

\section*{Acknowledgements}

 We would like to thank M. Morikawa and Y. Yasui for useful discussion and comments. We also thank H. Nishino and S. Rajpoot for bringing our attention to their papers.

\section*{Appendix}

Here we show some concrete calculations examining gravitational 
instantons are indeed the solutions of equation, (\ref{8Tnull2}) and 
(\ref{7Tnull2}). 
The first (\ref{8Tnull2}) for 8 dimension could be written in components as, 
\begin{eqnarray}
\zeta^a_{\mu \nu} \zeta^a_{\rho \sigma} \theta^{\rho \sigma} \w d \Psi = 0 
\end{eqnarray}
acting $\zeta^b_{\mu \nu}$ on both sides,
\begin{eqnarray}
\zeta^b_{\rho \sigma}\theta^{\rho \sigma}
 \w d\Psi = (2 \theta^{0b} - \varphi_{bij}\theta^{ij} ) \w d\Psi = 0 \label{8Tnull3}
\end{eqnarray}
Writing down (\ref{8Tnull3}) in components, 
\begin{eqnarray}
&&( -\theta^{01} + \theta^{23} + \theta^{65} + \theta^{47}) \w d\Psi = 0  \nonumber \\
&&( -\theta^{02} + \theta^{31} + \theta^{46} + \theta^{57}) \w d\Psi = 0  \nonumber \\
&&( -\theta^{03} + \theta^{67} + \theta^{12} + \theta^{54}) \w d\Psi = 0  \nonumber \\
&&( -\theta^{04} + \theta^{71} + \theta^{62} + \theta^{35}) \w d\Psi = 0  \nonumber \\
&&( -\theta^{05} + \theta^{43} + \theta^{16} + \theta^{72}) \w d\Psi = 0  \nonumber \\
&&( -\theta^{06} + \theta^{24} + \theta^{51} + \theta^{73}) \w d\Psi = 0  \nonumber \\
&&( -\theta^{07} + \theta^{36} + \theta^{25} + \theta^{14}) \w d\Psi = 0 .
\end{eqnarray}
Wedging appropriate $\theta^a$ to this relations, one could show that 
for all (a,b,c), $\theta^{abc} \w d\Psi = 0$, so $d\Psi = 0$. 
Proving for the second equation (\ref{7Tnull2}) of 7 dimension needs a little more patience. It could be written in components as,
\begin{eqnarray}
-\theta^a \w d \psi + \frac12 \varphi_{aij} \theta^{ij} \w d \varphi = 0 \label{7ins1}
\end{eqnarray}
wedge $\theta^b$ from left, 
\begin{eqnarray}
\frac13 \theta^{ab} \w d\psi + \frac12 \varphi_{aij} \theta^{bij}  \w d\varphi = 0 \label{7ins2}
\end{eqnarray}
act $\varphi_{abk}$,
\begin{eqnarray}
\frac13 \varphi_{abk} \theta^{ab} \w d\psi + \frac12 \psi_{bijk} \theta^{bij} \w d \varphi = 0 \label{7ins3}
\end{eqnarray} 
act $\varphi_{kcd}$ again,
\begin{eqnarray}
\frac13 \left\{ 2\theta^{cd} + \psi_{cdij} \theta^{ij} \right\} \w d\psi
+\frac32 \left\{ \varphi_{cij} \theta^{dij} - \varphi_{dij} \theta^{cij} \right\} \w d \varphi = 0 \label{7ins4}
\end{eqnarray}
with (\ref{7ins2}), (\ref{7ins4}) becomes,
\begin{eqnarray}
\left\{ \frac43 \theta^{cd} -  \psi_{cdij} \theta^{ij} \right\} \w d\psi = 0 \label{7ins5}
\end{eqnarray}
act $\varphi_{acd}$ on (\ref{7ins5}),
\begin{eqnarray}
\varphi_{aij} \theta^{ij} \w d\psi = 0 \label{7ins6}
\end{eqnarray}
then from (\ref{7ins3}),
\begin{eqnarray}
\psi_{aijk} \theta^{ijk} \w d\varphi = 0 \label{7ins7}
\end{eqnarray}
act $\psi_{klcd}$ on (\ref{7ins5}),
\begin{eqnarray}
\left\{ 4\theta^{kl} + \frac13 \psi_{klij} \theta^{ij} \right\} \w d\psi = 0 \label{7ins8}
\end{eqnarray} 
then from (\ref{7ins5}) and (\ref{7ins8}),
\begin{eqnarray}
\theta^{ij} \w d\psi = 0 \label{7ins9}
\end{eqnarray}
then we obtain, 
\begin{eqnarray}
d\psi = 0.
\end{eqnarray}
 From (\ref{7ins2}) and (\ref{7ins9}) 
\begin{eqnarray}
\varphi_{aij}\theta^{bij} \w d\varphi = 0 \label{7ins10}
\end{eqnarray}
and together with (\ref{7ins7}), 
we obtain 
\begin{eqnarray}
d\varphi = 0 
\end{eqnarray}
by similar calculation as in 8-dimensional case.


\end{document}